\documentclass[twocolumn,showpacs,preprintnumbers,amsmath,amssymb]{revtex4}

\def\undersim#1{\setbox9\hbox{${#1}$}{#1}\kern-\wd9\lower
    2.5pt \hbox{\lower\dp9\hbox to \wd9{\hss $_\sim$\hss}}}
\usepackage{amssymb}
\usepackage{amsmath}
\usepackage{graphicx}
\usepackage{color}

\def\undersim#1{\setbox9\hbox{${#1}$}{#1}\kern-\wd9\lower
    2.5pt \hbox{\lower\dp9\hbox to \wd9{\hss $_\sim$\hss}}}

\def\mk{{\mathbf p}}

\begin{document}

\title{Squeezed back-to-back correlation between bosons and antibosons with different
in-medium masses in high-energy heavy-ion collisions}

\author{Peng-Zhi Xu$^1$}
\author{Wei-Ning Zhang$^{1,2}$\footnote{wnzhang@hit.edu.cn, wnzhang@dlut.edu.cn}}
\author{Yong Zhang$^3$}
\affiliation{$^1$Department of Physics, Harbin Institute of Technology, Harbin,
Heilongjiang 150006, China\\
$^2$School of Physics, Dalian University of Technology, Dalian, Liaoning 116024,
China\\
$^3$School of Mathematics and Physics, Jiangsu University of Technology, Changzhou,
Jiangsu 213001, China}


\begin{abstract}
We derive the formulas of the squeezed back-to-back correlation (SBBC) between a boson
and antiboson with different in-medium masses in high-energy heavy-ion collisions.
The influence of the in-medium mass difference between a boson and antiboson on the
SBBC is investigated.  We calculate the SBBC functions of $D$-meson pairs for the
hydrodynamic sources described by the VISH2+1 code for Au+Au collisions at $\sqrt{s_{NN}}
=200$~GeV.  Our results indicate that the SBBC strengths of $D^+D^-$ and
$D^0{\bar D}^0$ are different if there are charge-dependent in-medium
interactions.
\end{abstract}

\maketitle

\section{Introduction}
In high-energy heavy-ion collisions, the in-medium mass shifts of bosons may cause
a squeezed back-to-back correlation (SBBC) between a detected boson and antiboson
\cite{AsaCso96,AsaCsoGyu99,Padula06,Zhang15a}.  This SBBC is related to the in-medium
energies of the bosons, through a Bogoliubov transformation between the creation
(annihilation) operators of the quasiparticles in the medium and the corresponding free
particles \cite{AsaCso96,AsaCsoGyu99,Padula06,Zhang15a}.  The study of the SBBC can
provide information about boson formations and in-medium interactions in
high-energy heavy-ion collisions.

In previous studies of the SBBC, the mass shifts of a boson and antiboson are taken
to be the same \cite{{AsaCso96,AsaCsoGyu99,Padula06,Zhang15a,Zhang15,Zhang-IJMPE15,
Zhang-EPJC16,DudePadu10,PaduSoco10}}.  More generally, the interactions of a boson
and antiboson in a medium are different, especially in a medium with a finite
baryon chemical potential \cite{{LiLeeBrown-NPA97,SibTsuTho-EPJA99,CMKo-JPG01,
Mishra-PRC69-04,Mishra-PRC70-04,BlaCosKal-PRD12}}.  The in-medium energy difference
between a boson and an antiboson leads to a mass difference between the quasiparticles
in a medium.  It is necessary to check the validity of the previous formulas of the 
SBBC calculations in this case.

In this work, we derive the formulas for calculating the SBBC function of a
boson-antiboson with different in-medium masses.  The influence of the in-medium
mass difference on the SBBC functions of $D$-meson pairs is investigated.
Since $D$-mesons contain a charm quark, which is believed to experience the entire
evolution of the quark-gluon plasma (QGP) created in relativistic heavy-ion
collisions, $D$-meson measurements have recently attracted great interest
\cite{STAR-PRL14,STAR-NPA16,ALICE-PRL13,ALICE-PRC14,ALICE-JHEP16,ALICE-PRL18}.
We calculate the SBBC functions of $D$-meson pairs for the hydrodynamic sources
described by the VISH2+1 code \cite{VISH2+1} and find that the SBBC strengths
of $D^+D^-$ and $D^0{\bar D}^0$ are different in Au+Au collisions at
$\sqrt{s_{NN}}=200$~GeV if there are charge-dependent in-medium interactions.

In Section II, we present the formula derivations of the SBBC function for
a boson and antiboson with different in-medium masses.  Then, we show
the SBBC results of $D$-meson pairs in Section III.  Finally, a
summary is given in Section IV.

\section{Formulas}
In the framework of a complex scalar field, the Hamiltonian density of a system
for a free boson and antiboson with mass $m$ is given by:
\begin{equation}
{\cal H}={\dot\phi}{\dot\phi^\dag}+{\nabla\phi^\dag}\cdot{\nabla\phi} +m^2
\phi^\dag \phi,
\end{equation}
where
\begin{equation}
\phi(x)=\sum_{\mk}(2V\omega_\mk)^{-\frac{1}{2}}\left(e^{-ip\cdot x}a_\mk
+e^{ip\cdot x}b^\dag_\mk \, \right),
\end{equation}
\begin{equation}
\phi^\dag(x)=\sum_{\mk}(2V\omega_\mk)^{-\frac{1}{2}}\left(e^{ip\cdot x}
a^\dag_\mk +e^{-ip\cdot x}b_\mk \, \right),
\end{equation}
where $a_\mk$ and $a^\dag_\mk$ ($b_\mk$ and $b^\dag_\mk$) are creation and
annihilation operators of the free boson (antiboson), respectively, $p=(\omega_\mk, \mk)$,
and $\omega_\mk=\sqrt{\mk^2+m^2}$.

For a boson and antiboson in a medium with the same mass $m'=\sqrt{m^2\pm m_1^2}$,
where ``$+$" or ``$-$" represents the case that $m'$ is larger or smaller than
$m$, respectively, the Hamiltonian density of system can be written as \cite{AsaCso96}
\begin{equation}
{\cal H}_{\rm M}={\dot\phi}{\dot\phi^\dag}+{\nabla\phi^\dag}\cdot{\nabla\phi}
+(m^2\pm m_1^2) \phi^\dag \phi,
\end{equation}
and the Hamiltonian of the system can be diagonalized through a Bogoliubov
transformation \cite{AsaCso96,AsaCsoGyu99}.

Generally speaking, the interactions of a boson and antiboson with a medium are
somewhat different.  Assuming the energy split between the boson and antiboson
in the medium is $2\delta'$, we consider the transformation
\begin{equation}
\phi \to e^{i\delta't}\phi, ~~~~\phi^\dag \to e^{-i\delta't}\phi^\dag,
\end{equation}
and have
\begin{eqnarray}
{\cal H}_{\rm M}&=&{\dot \phi}{\dot \phi}^\dag +{\nabla\phi^\dag}\cdot{\nabla
\phi} +m^2 \phi^\dag \phi \pm m_1^2 \phi^\dag \phi \nonumber\\
&&+\delta'^2\phi\phi^\dag -i\delta'({\dot\phi}\phi^\dag-\phi{\dot\phi}^\dag).
\end{eqnarray}
It will be seen that $\delta'^2$ provides an additional term of mass square in
average energy of the boson and antiboson, which is associated with the different
in-medium interactions for both, while $m_1^2$ reflects the
in-medium interaction that is the same for both.

Using the Bogoliubov transformation between the operators $(a_\mk, a^\dag_\mk,
b_\mk, b^\dag_\mk)$ for the free particles and $(a'_\mk, a'^\dag_\mk, b'_\mk,
b'^\dag_\mk)$ for the quasiparticles,
\begin{equation}
a_\mk =c_\mk a'_\mk +s^*_{-\mk} b'^\dag_{-\mk}, ~~~~
b_\mk ={\bar c}_\mk b'_\mk +{\bar s}^*_{-\mk} a'^\dag_{-\mk},
\end{equation}
where
\begin{equation}
c^*_{\pm\mk}=c_{\pm\mk}={\bar c}^*_{\pm\mk}={\bar c}_{\pm\mk}=\cosh r_\mk,
\end{equation}
\begin{equation}
s^*_{\pm\mk}=s_{\pm\mk}={\bar s}^*_{\pm\mk}={\bar s}_{\pm\mk}=\sinh r_\mk,
\end{equation}
\begin{equation}
r_\mk =\frac{1}{2} \ln\left({\omega_\mk}/{\Omega_\mk}\right),
\end{equation}
\begin{equation}
\Omega_\mk =\sqrt{\mk^2 +m^2 \pm m_1^2 +\delta'^2},
\end{equation}
we can diagonalize the Hamiltonian of the system for the boson and antiboson with an
energy split $2\delta'$ in the medium as:
\begin{equation}
H_{\rm M}=\sum_\mk \left[(\Omega_\mk +\delta')a'^\dag_\mk a'_\mk+(\Omega_\mk
-\delta')b'^\dag_\mk b'_\mk \right].
\end{equation}
The in-medium masses of the boson and antiboson are:
\begin{equation}
m'_{\pm}= (\Omega_{\mk} \pm \delta')\big|_{\mk=0}=\sqrt{m^2 \pm m_1^2 +\delta'^2}
\pm \delta',
\end{equation}
and the in-medium mass difference between the boson and antiboson with the same momentum
is also the split $2\delta'$.

It can be seen that the Bogoliubov transformation involves only the average
in-medium energy of the boson and antiboson.  Furthermore, the average energy $\Omega_\mk$
is not only related to $m_1$ associated with the in-medium interactions that are the
same for the boson and antiboson, but it is also related to the $\delta'$ associated with
the in-medium interactions that are different for the boson and antiboson.

The SBBC function of the boson-antiboson with momenta $\mk_1$ and $\mk_2$ is
defined as: \cite{AsaCsoGyu99,Padula06}
\begin{equation}
\label{BBCf}
C(\mk_1,\mk_2) = 1 + \frac{|G_s(\mk_1,\mk_2)|^2}{G_c(\mk_1,\mk_1) G_c(\mk_2,
\mk_2)},
\end{equation}
where $G_c(\mk_1,\mk_2)$ and $G_s(\mk_1,\mk_2)$ are the so-called chaotic and
squeezed amplitudes \cite{AsaCsoGyu99,Padula06}, respectively.  They are given by:
\cite{AsaCsoGyu99,Padula06,MakhSiny,Zhang15a}
\begin{eqnarray}
\label{Gchydro}
&&\hspace*{-3mm}G_c({\mk_1},{\mk_2})\!=\!\int \frac{d^4\sigma_{\mu}(r)}{(2\pi)^3}
K^\mu_{1,2}\, e^{i\,q_{1,2}\cdot r}\,\! \Bigl\{|c'_{\mk'_1,\mk'_2}|^2\,
n'_{\mk'_1,\mk'_2}~~\nonumber \\
&& \hspace*{16mm}
+\,|s'_{-\mk'_1,-\mk'_2}|^2\,[\,n'_{-\mk'_1,-\mk'_2}+1]\Bigr\},\\[1ex]
\label{Gshydro}
&&\hspace*{-3mm}G_s({\mk_1},{\mk_2})\!=\!\int \frac{d^4\sigma_{\mu}(r)}{(2\pi)^3}
K^\mu_{1,2}\, e^{2 i\,K_{1,2}\cdot r}\!\Bigl\{s'^*_{-\mk'_1,\mk'_2}
c'_{\mk'_2,-\mk'_1}~~\nonumber \\
&& \hspace*{16mm}
\times n'_{-\mk'_1,\mk'_2}+c'_{\mk'_1,-\mk'_2} s'^*_{-\mk'_2,\mk'_1}
[n'_{\mk'_1,-\mk'_2} + 1] \Bigr\},\nonumber\\
\end{eqnarray}
for an evolving source.  Here, $d^4\sigma_{\mu}(r)$ is the four-dimensional
element of the freeze-out hyper-surface, $q^{\mu}_{1,2}=p^{\mu}_1-p^{\mu}_2$,
$K^{\mu}_{1,2} =(p^{\mu}_1+p^{\mu}_2)/2$, and $\mk_i'$ is the local-frame momentum
corresponding to $\mk_i ~(i=1,2)$.  In Eqs.\ (\ref{Gchydro}) and (\ref{Gshydro}),
the quantities $c'_{\mk'_1,\mk'_2}$ and $s'_{\mk'_1,\mk'_2}$ are the coefficients
of the Bogoliubov transformation between the creation (annihilation) operators of the
quasiparticles and the free particles, respectively, and $n'_{\mk'_1,\mk'_2}$ is the boson
distribution of the quasiparticle pair \cite{AsaCsoGyu99,Padula06,Zhang15a,Zhang15}.

\section{Results}
We first consider a simple case, namely a rest particle-emitting source with a fixed
freeze-out temperature $T_f$, a Gaussian spatial distribution $[e^{-r^2/2R^2}/
(\sqrt{2\pi}R)^3]$, and a temporal distribution of exponential decay $[\theta(t-t_0)
e^{-(t-t_0)/\Delta t}/\Delta t]$ \cite{AsaCsoGyu99,Padula06,Zhang15}.  In this case,
the SBBC function of the boson-antiboson emitted from the source with momenta $\mk_1$ and
$\mk_2$, and under the condition $|\mk_1|=|\mk_2|=|\mk|$, can be given analytically
by \cite{Yang-CPL18}
\begin{eqnarray}
C(\mk_1,\mk_2)&=&1+e^{-2\mk^2\!R^2[1+\cos(\alpha)]}B(\mk)\cr
&\equiv& 1+f(\alpha)B(\mk),
\end{eqnarray}
where $\alpha~(0<\alpha<\pi)$ is the angle between momenta $\mk_1$ and $\mk_2$,
and
\begin{eqnarray}
\label{Bp}
B(\mk)=\frac{|c_{\mk}\,s_{\mk}^*\,n_{\mk} +c_{\mk}s_{\mk}^*\,(n_{\mk}+1)|^2}
{(1+4\omega_\mk^2\Delta t^2)\,n_1(\mk)\,n_1(\mk)},
\end{eqnarray}
where $n_\mk$ is the boson distribution of the quasiparticle pair with average energy
$\Omega_\mk$, and $n_1(\mk)=|c_{\mk}|^2\,n_{\mk}+|s_{\mk}|^2 (n_{\mk}+1)$.
Here, it should be mentioned that we have used an approximation that replaces
the boson or antiboson momentum distribution with the pair momentum distribution
$n_\mk$ in the denominator of Eq.~(\ref{Bp}).
The SBBC function $C(\mk_1,\mk_2)$ approaches its maximum $[1+B(\mk)]$ when the
boson and antiboson approach antiparallelism, and decreases with increasing $\cos\alpha$
exponentiality.  For the case of incomplete antiparallelism $\mk_1$ and $\mk_2$, the mass-shift-caused
SBBC still exists, except for very large sources.

\begin{figure}[htbp]
\includegraphics[scale=0.67]{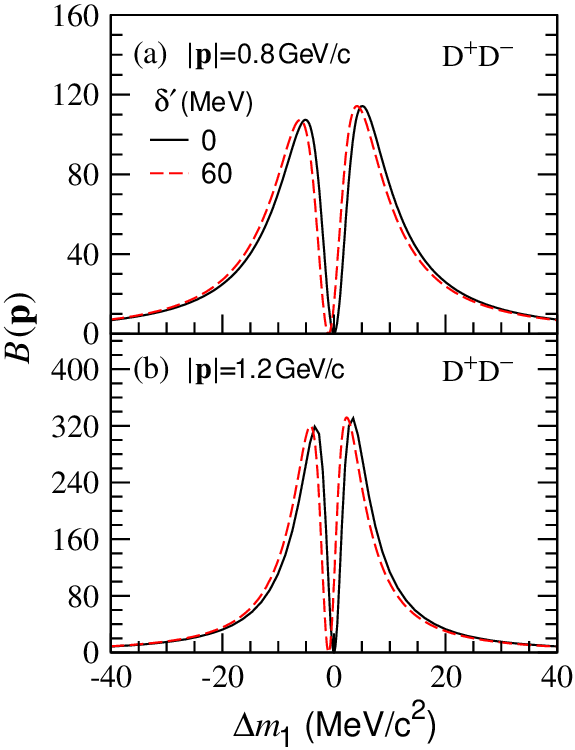}
\caption{(Color online) $B(\mk)$ as functions of mass shift $\Delta m_1=(m'-m_0)$
for $D^+D^-$ pairs with momenta 0.8 and 1.2 GeV/$c$ and the splits $\delta'=0$
and 60 MeV (solid and dashed lines). }
\label{fz1}
\end{figure}

\begin{figure}[htbp]
\includegraphics[scale=0.67]{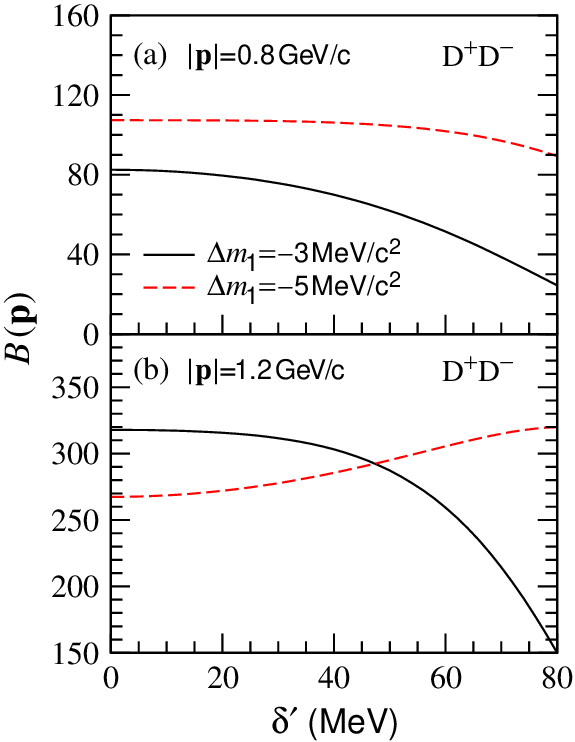}
\caption{(Color online) $B(\mk)$ as functions of $\delta'$ for $D^+D^-$ pairs
with momenta 0.8 and 1.2 GeV/$c$ and $\Delta m_1=-3$ and $-5$~MeV/$c$. }
\label{fz11}
\end{figure}

The SBBC is expected to be strong for the mesons with large masses
\cite{Zhang-EPJC16,Padula-JPG10}, under the same source size and freeze-out
temperature.  We plot $B(\mk)$ in Fig.~\ref{fz1} as functions of mass shift
$\Delta m_1=(m'-m_0)$ for $D^+D^-$ pairs with different momenta and in-medium
mass differences.  Here, the solid and dashed lines are for $\delta'=$ 0 and 60
MeV, respectively.  In the calculations, the source freeze-out temperature is
taken to be 150 MeV and we take $\Delta t=2$ fm.  It can be seen that $\delta'$
leads to a shift of $B(\mk)$ towards decreasing $\Delta m_1$.  The function
width decreases with increasing momentum.

In Fig.~\ref{fz11} we plot $B(\mk)$ as functions of $\delta'$ for $D^+D^-$
pairs with momenta 0.8 and 1.2 GeV/$c$.  Here, the source parameters are the 
same as in Fig.~\ref{fz1}.  Based on the results \cite{Yang-CPL18} calculated 
in the FMFK framework \cite{FMFK-PRC06,MFFK-PRL04}, the mass of the $D$-meson 
in a hadronic medium in relativistic heavy-ion collisions is approximately $3\sim
5$~MeV/$c^2$ smaller than its value at a free state.  Thus, we compare the $B(\mk)$
functions at $\Delta m_1=-3$ and $-5$ MeV/$c^2$.  For the lower momentum $|\mk|
=0.8$~GeV/$c$, $B(\mk)$ decreases with increasing $\delta'$.  However, for the
higher momentum $\mk=1.2$~GeV/$c$, the result of $B(\mk)$ for $\Delta m_1=-5$
MeV/$c^2$ increases with increasing $\delta'$, while the result for $\Delta m_1=
-3$ decreases more rapidly with increasing $\delta'$ when $\delta'>40$~MeV.
The results of $B(\mk)$ are sensitive to the mass shift $\Delta m_1$, mass
difference $\delta'$, and particle momentum $|\mk|$.

\begin{figure}[htbp]
\includegraphics[scale=0.70]{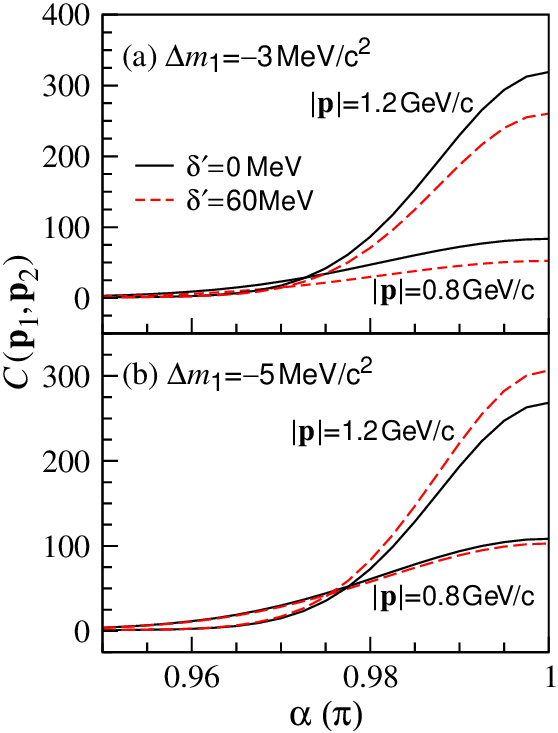}
\caption{(Color online) SBBC functions of $D$-meson pairs for the Gaussian source as
in Figs.~\ref{fz1} and \ref{fz11} and with $R=3$ fm. }
\label{fz2}
\end{figure}

We show in Fig.~\ref{fz2} the SBBC functions of $D$-meson pairs for the source as in
Figs.~\ref{fz1} and \ref{fz11} and with a Gaussian radius $R=3$ fm.  It can be seen
that the influences of $\delta'$ on the SBBC functions at the higher momentum are
different when $\Delta m_1=-3$ and $-5$~MeV/$c^2$.  For $\Delta m_1=-3$~MeV/$c^2$,
$\delta'$ makes the SBBC function at high momentum decrease.  However, $\delta'$
makes the SBBC function at high momentum increase for $\Delta m_1=-5$~MeV/$c^2$.

As discussed above, $\delta'$ is associated with the in-medium interactions, which
are different for a boson and antiboson.  If assuming these in-medium interactions
are particle-charge-dependent, the split $\delta'$ will be zero for a
$D^0{\bar D}^0$ pair.  For a $D^+D^-$ pair, the split may reach a few tens
of MeV \cite{SibTsuTho-EPJA99,BlaCosKal-PRD12}.  In this case, it can be seen that
the SBBC of $D^+D^-$ at high momentum is weaker or stronger than that of
$D^0{\bar D}^0$ when $\Delta m_1=-3$~MeV/$c^2$ or $\Delta m_1=-5$~MeV/$c^2$
in the simple source model, where $\Delta m_1$ is the mass-shift related to the
in-medium interactions, which are the same for the particles and antiparticles.

\begin{figure}[htbp]
\includegraphics[scale=0.6]{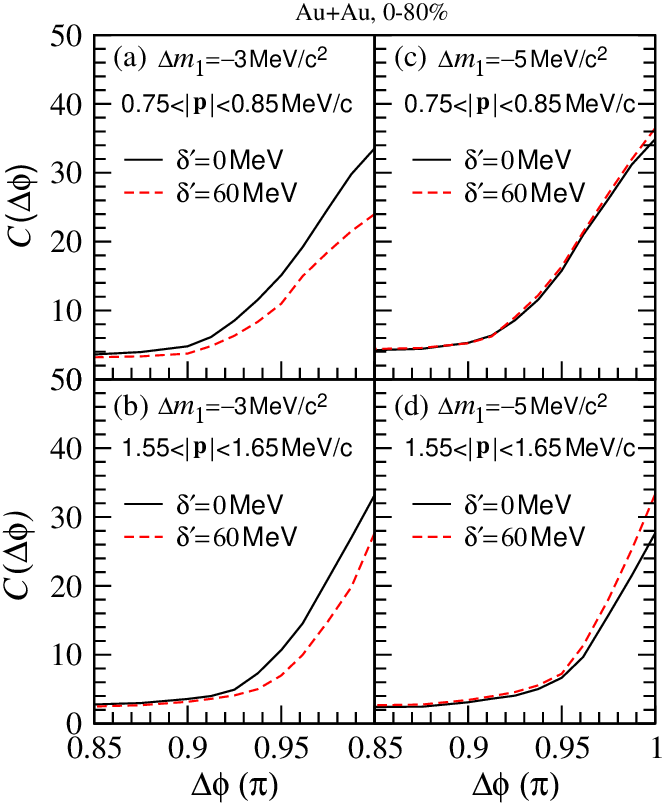}
\caption{(Color online) SBBC functions of $D$-meson pairs for viscous hydrodynamic
sources for $\sqrt{s_{NN}}=200$~GeV Au+Au collisions with 0\%--80\% centrality. }
\label{fz3D-VISH08}
\end{figure}

\begin{figure}[htbp]
\includegraphics[scale=0.6]{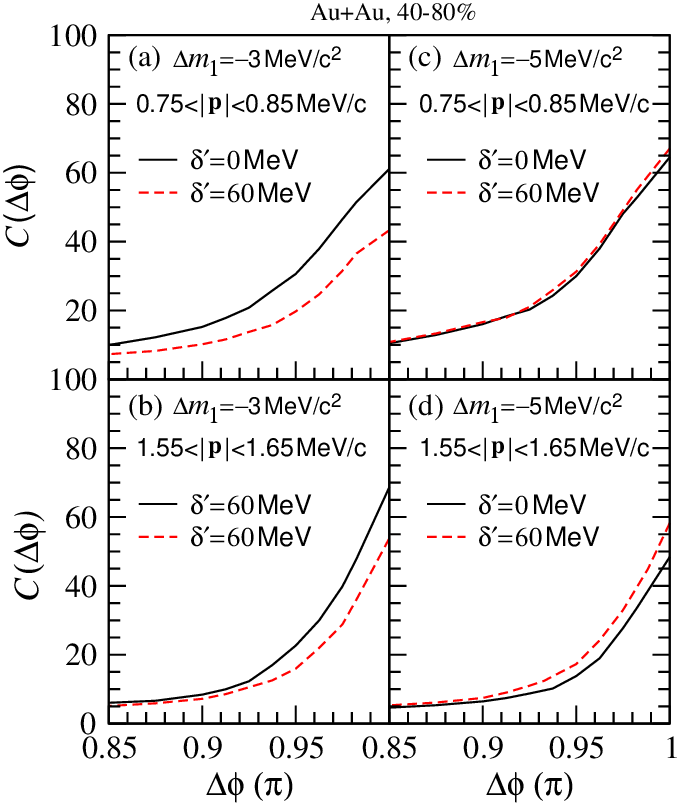}
\caption{(Color online) Same as Fig.~\ref{fz3D-VISH08}, but with 40\%--80\% centrality. }
\label{fz3D-VISH48}
\end{figure}

We next investigate the SBBC functions for the evolving sources described by the
viscous hydrodynamic model VISH2+1 \cite{VISH2+1} under the MC-Glb initial conditions
fluctuating event-by-event \cite{VISHb}.  Figures \ref{fz3D-VISH08} and \ref{fz3D-VISH48}
show the SBBC functions $C(\Delta \phi)$ of $D$-meson pairs for the hydrodynamic sources
for $\sqrt{s_{NN}}=200$~GeV Au+Au collisions with centralities 0\%--80\% and 40\%--80\%,
respectively.  Here, $\Delta \phi$ is the angle between the transverse momenta of the two
$D$ mesons, the ratio of the shear viscosity to entropy density of QGP is taken to be 0.08
\cite{Shen11-prc,Qian16-prc}, and we take the freeze-out temperature to be 150~MeV based
on comparisons of the transverse-momentum spectra of $D$ mesons \cite{Yang-CPL18} with the
RHIC experimental data \cite{STAR-PRL14}.

It can be seen that the results of the SBBC function for nonzero $\delta'$ are smaller than those
for zero $\delta'$ when $\Delta m_1=-3$~MeV.  In addition, the results of the SBBC function for nonzero
$\delta'$ are slightly larger than those for zero $\delta'$ when $\Delta m_1=-5$~MeV.  Assuming that
there the split $\delta'$ associated with the charge-depend in-medium
interactions exists, we conclude that the SBBC of $D^+D^-$ and $D^0{\bar D}^0$ pairs are
different for the different $\Delta m_1$.  This may provide a probe with which to study the in-medium
interactions in detail.

The dependence of SBBC on particle momentum is complicated for the hydrodynamic sources
with fluctuating initial conditions.  The more serious oscillations of single-event SBBC
functions at higher momentum \cite{Zhang15a} may lead to a lower SBBC function after
being averaged over events \cite{Zhang-EPJC16,Yang-CPL18}, although the intercept of the
SBBC function $C(\mk,-\mk)$ increases with increasing particle momentum \cite{{AsaCso96,
AsaCsoGyu99,Padula06,Zhang15a,Zhang15,Zhang-IJMPE15,Zhang-EPJC16,DudePadu10,PaduSoco10}}.
It can be seen that the widths of the SBBC functions $C(\Delta\phi)$ for higher momentum
are narrower than those for lower momentum, which is similar to that for the simple
source in Fig.~\ref{fz2}.  By comparing the SBBC functions in Figs.~\ref{fz3D-VISH08}
and \ref{fz3D-VISH48}, we find that the SBBC functions for the peripheral collisions are
higher than those for the central collisions.  This is because the averaged source lifetime
is smaller for the peripheral collisions \cite{Zhang-EPJC16}.

\section{Summary}
We derived the formulas of SBBC between a boson and antiboson with different
in-medium masses.  The SBBC is related to the average in-medium energy of the boson
and antiboson, which is the same as for quasiparticles having the same mass.
However, the more general formulas developed in this paper indicate that the SBBC is
associated with both in-medium interactions, one is the same for a boson and antiboson and the 
other one is different for the boson and antiboson, respectively.  Due to the high strength, the
SBBC of heavy-meson pairs provide a possible probe with which to study the in-medium interactions
of the heavy mesons in detail in relativistic heavy-ion collisions.  Our results
calculated with the VISH2+1 code for Au+Au collisions at $\sqrt{s_{NN}}=200$~GeV
indicate that the SBBC strengths of $D^+D^-$ and $D^0{\bar D}^0$ pairs are
different if there are charge-dependent in-medium interactions.

\begin{acknowledgments}
This research was supported by the National Natural Science Foundation of China
under Grant Nos. 11675034 and 11647166.
\end{acknowledgments}

\end{document}